\documentstyle[preprint,aps]{revtex}
\begin{document}
\baselineskip=15pt
\title{Phase diagrams of electronic state \\
 on   One  Dimensional  d-p Model}
\author{
Kazuhiro SANO   ~and~  Yoshiaki \={O}NO\raisebox{0.5ex}{*} \\
 Department of Physics, Faculty of Education, Mie University, \\
  Tsu,  Mie  514.       \\ \raisebox{0.5ex}{*}Department  of  Physics,
Nagoya University, Nagoya 464-01}
\maketitle
\begin{abstract}
   We investigate the one-dimensional(1D) d-p model, simulating a Cu-O
linear  chain with strong  Coulomb repulsion,  by using the  numerical
diagonalization method.
Using the Luttinger liquid theory, we obtained  phase diagrams of  the
ground state on $U_d-U_{pd}$ plane, where $U_d$ and $U_{pd}$ represent
on-site  interaction at d-sites and  the nearest-neighbor  interaction
between p- and d-sites respectively.
In  the  weak  coupling  region,  they  agree  with  the  g-ology;   a
superconducting phase (SC(I)) is restricted to attractive  interaction
$U_{pd}<0$. On the other hand, in the strong coupling region, we found
a  novel  superconducting  phase(SC(II))   for  repulsive  interaction
$U_{pd}>|U_d|$   and   a  insulating  state  with  a  charge  gap  for
$U_d>U_d^c$  and  $U_{pd}>U_{pd}^c$ with critical values  $U_d^c$  and
$U_{pd}^c$ at half-filling.
 Away  from  half-filling,  another  superconducting  phase  (SC(III))
appears  for  $U_d>>U_{pd}>0$;  which  has  been  found  for   $U_d\to
\infty$  in  the previous paper [Physica {\bf C205} (1993)  170].   An
analysis of the spin gap suggests that  the SC(I) and SC(II)   include
the  Luther-Emery  region (with spin-gap) in part, while  the  SC(III)
belongs to Tomonaga-Luttinger region (without spin-gap) in whole.

 \end{abstract}

\newpage
 \section{Introduction}
 Since  the  discovery  of copper  oxides   superconductors,  strongly
correlated   electron  systems  have  been  extensively  studied.   In
particular, there is much theoretical interest in the electronic state
of  the   d-p  model  because  of  the  possible  relevance  of  high-
temperature superconductivity[1-12].
 In  the  previous work, the present authors have  studied  the   one-
dimensional  d-p model, simulating a Cu-O linear chain  with  infinite
intra-site  interaction  $U_d$at  the  Cu-(d-)sites  and  the  nearest
neighbor   interaction $U_{pd}$ between the O-(p-)sites  and  d-sites,
is   solved   exactly   by   using   the   numerical   diagonalization
method\cite{Sanoono1}.
   By  assuming  the Luttinger liquid relations,  the  superconducting
correlation  is  found to be  dominant compared with the CDW  and  SDW
correlations  in the proximity of the phase boundary towards the phase
separation.
 Recently,  A.Sudb$\phi$ et al. showed that the  d-p chain with  large
$U_{pd}$   exhibits  flux  quantization  with  charge  $2e$  and  slow
algebraic decay of the singlet superconducting correlation function on
oxygen sites\cite{Sudbs}.
   These  works   demonstrate  that   the  numerical   diagonalization
studies of  finite sizes systems combined the Luttinger liquid  theory
have supplied us with unambiguous and important information about  the
complicated     electronic     systems     such     as     the     d-p
model\cite{Sanoono1,Sudbs,Dagotto,Mila,Sanoono2}.

   In  this  paper, we study  the  one-dimensional d-p  model  by  the
numerical  diagonalization method. To clarify the role of  the  inter-
and intra-site interaction  for the electronic structure of d-p model,
we turn our attention to  $U_d$ and $U_{pd}$.
In  the  weak  coupling regime, we will take  advantage  of  the  weak
coupling theory (so-called g-ology) to analyze  numerical results. For
$U_{pd}  \to  \infty$, we find some exact results  which   agree  with
numerical results.
This paper is organized as follows.
In  the next section, we define the model Hamiltonian. The   Luttinger
liquid  relation   is  also  discussed.  In  Sec.3,   we  present  our
numerical results of phase diagram on the $U_d-U_{pd}$ plane by  using
the  Luttinger  Liquid  relation.  We also analyze  a  spin  gap.  The
conclusion of this work is presented in Sec. 4.

\section{  Model and Luttinger liquid relation}
We consider the following model Hamiltonian for the  Cu-O chain:

\begin{eqnarray}
  H&=&-t\sum_{<ij>,\sigma} (p_{i\sigma}^{\dagger} d_{j\sigma}+h.c.)
    +\epsilon_{d}\sum_{j,\sigma}      d_{j\sigma}^{\dagger}d_{j\sigma}
+\epsilon_{p}\sum_{i,\sigma}          p_{i\sigma}^{\dagger}p_{i\sigma}
\nonumber \\
   &+& U_d\sum_{j}n_{dj\uparrow}n_{dj\downarrow}
    +   U_{pd}\sum_{<ij>,\sigma   \sigma'   }n_{pi\sigma}n_{dj\sigma'}
\quad ,
\end{eqnarray}
where  $d^{\dagger}_{j\sigma}$ and $p^{\dagger}_{i\sigma}$  stand  for
creation operators of a electron  with spin $\sigma$ in  the d-orbital
at  site $j$ and of a electron with spin $\sigma$ in the p-orbital  at
site $i$, respectively.
   $n_{dj\sigma}=d_{j\sigma}^{\dagger}d_{j\sigma}$                 and
$n_{pi\sigma}=p_{i\sigma}^{\dagger}p_{i\sigma}$.
   $t$ stands for the transfer energy  between the nearest neighbor d-
and  p-sites, which will be set to be unity ($t$=1) hereafter  in  the
present study.
 The   atomic  energy   levels  of p-  and  d-orbitals  are  given  by
$\epsilon_{p}$  and $\epsilon_{d}$, respectively. The  charge-transfer
energy $\Delta$ is defined as $\Delta=\epsilon_{p}-\epsilon_{d}$.

  The fillingness $n$ is defined  by  $n=N_{e}/N_{u}$, where $N_u$  is
the total number of unit cells (each unit cell contains a d- and a  p-
orbital),   $N_e$  is  the total electron number and  the  Fermi  wave
number  $k_F$ is given as $k_F=\frac{\pi }{2}n$.
  In  the  non-interacting limit with $U_d=U_{pd}=0$,  the  hybridized
bands are given as

$$    E^{\pm}(k)=    \frac{\epsilon_{p}+\epsilon_{d}    \pm     \sqrt{
\Delta^2+4t_k^2} }{2},   \eqno{(2)}$$
   where      $t_k=2tcos(k/2)$     and      $k=2\pi      \ell/N_u\quad
(\ell=0,\pm1,\pm2,...)$.  Here, $E^+(k)$ and $E^-(k)$ stand p- and  d-
like bands respectively.

  To   achieve  systematic calculation, we use the  periodic  boundary
condition  for  $N_e=4m+2$  and antiperiodic  boundary  condition  for
$N_e=4m$ with  an  integer $m$. This choice of the boundary  condition
removes  accidental degeneracies so that the ground state is always  a
singlet with zero momentum.
 We numerically diagonalize the Hamiltonian with up to 12 sites (6unit
cells)  using the standard Lanczos algorithm.
  We  have  calculated  the ground state energies per  unit  cell  for
different sizes of systems.
  The  relative  difference is of order  of  $10^{-3}\sim10^{-4}$  for
typical values of parameters involved in the model.
   It  indicates  that   size dependence in the ground state  energies
is  negligible.

   The chemical potential $\mu (N_{e},N_{u})$ is defined by
$$
   \mu             (N_{e},N_{u})=\frac{E_{0}(N_{e}+1,N_u)-E_{0}(N_{e}-
1,N_u)}2\quad , \eqno{(3)}
$$
where  $E_{0}(N_e,N_u)$ is the  ground state energy of a  system  with
$N_u$ unit cells and $N_e$ electrons.
     When  the  charge gap vanishes in the  thermodynamic  limit,  the
uniform  charge susceptibility $\chi_c$ is obtained from
$$
 \chi_c(N_{e},N_{u})=\frac{2/N_u}{\mu(N_{e}+1,N_{u})-\mu(N_{e}-
1,N_{u})}  \quad .\eqno{(4)}
$$

    We will discuss the  correlation functions  in connection with the
Luttinger liquid theory\cite{Haldane,Haldane2,Ogata2}.
  Some  relations  obtained in the Luttinger liquid theory  have  been
established as universal relations in various one-dimensional  models.
Some  of  one-dimensional  models  can be solved  rigorously  by   the
combined  use  of the Bethe ansatz method with the  numerical  methods
and/or the  conformal filed theory\cite{Ogatashiba,Schulz,Kawakami}.
  In these theories, the critical exponents describing  the  power-law
decay of various types of correlation functions have  been determined.
  The     bosonization    theory    on     the      Tomonaga-Luttinger
model\cite{Tomonaga,Luttinger}        and       the       Luther-Emery
model\cite{Luther} and the week  coupling renormalization group theory
(known    as   g-ology)   also   provides   us  with   some   rigorous
information about the critical  exponents\cite{Kimura,Solyom}.
    In  the  Luttinger  liquid  theory,   the  critical  exponents  of
correlation  functions are determined by a single  parameter  $K_\rho$
for  isotropic  models  in  spin  space,  or  more  explicitly,   some
correlation functions in momentum space have singularities as follows:

\quad

1) Tomonaga-Luttinger (T-L)  regime (without spin-gap)

$$
   C_{CDW}(k)\quad  \mbox{ and }\quad  C_{SDW}(k)\sim|k-2k_F|^{K_\rho}
\qquad\mbox{  for}\qquad k\sim 2k_F \quad ,   \eqno{(5.a)}
$$

$$
        C_{SS}(k)\sim\left\{   \begin{array}{l}   |k|^{\frac1{K_\rho}}
\qquad\qquad\qquad\quad \;  \mbox{  for} \qquad \; k\sim 0 \quad,
       \\|k-2k_F|^{\frac1{K_\rho}+K_{\rho}-1}
        \qquad\mbox{  for}  \qquad k\sim 2k_F \quad ,
\end{array}\right.     \eqno{(5.b)}
$$
$$
    C_{TS}(k)\sim|k|^{\frac1{K_\rho}}    \qquad\mbox{    for}   \qquad
k\sim0 \quad , \eqno{(5.b)}
$$

2) Luther-Emery (L-E)  regime (with spin-gap)

$$
   C_{CDW}(k)\sim|k-2k_F|^{K_\rho-1}  \qquad\mbox{   for}\qquad  k\sim
2k_F \quad ,   \eqno{(6.a)}
$$

$$
        C_{SS}(k)\sim |k|^{\frac1{K_\rho}-1}   \qquad\qquad\qquad\quad
\mbox{  for} \qquad  k\sim 0 \quad , \qquad
     \eqno{(6.b)}
$$
$$
   C_{SDW}(k)   \quad\mbox{  and }\quad   C_{TS}(k)   \qquad\mbox{  no
singularity }  \quad , \eqno{(6.c)}
$$
where  $ C_{CDW}$, $ C_{SDW}$, $C_{SS}$ and $C_{TS}$ stand for  Charge
Density  Wave (CDW), Spin Density Wave (SDW), Singlet  Superconducting
and Triplet Superconducting correlation functions respectively.
In  the Luther-Emery regime,  the spin excitation spectrum has a  gap,
while in the Tomonaga-Luttinger regime, the spin  is gapless.
In the T-L regime,
spin excitation.
the SS and TS correlation have the same critical exponent apart from a
logarithmic correction.
On  the  other hand, in the L-E regime,  the  real  space  correlation
functions of SDW and TS decrease exponentially with distance $r$.
The  parameter  $K_{\rho}$  is related to  the  charge  susceptibility
$\chi_c$     and     the    charge    velocity    $v_c$     by     the
relations\cite{Haldane,Ogata2,Schulz},
$$
          K\sb{\rho}=\frac{\pi}{2}v\sb{c}\chi\sb{c}      \quad       ,
\eqno{(7)}
$$
$$
           v_c=\frac{N_u}{2\pi}(E_{1}-E_{0}) \quad ,  \eqno{(8)}
$$
where $E_{1}-E_{0}$ is the lowest charge excitation energy.
    These relations  tell us that the  enhancement  of  $\chi_c$ leads
to  increase  of  absolute  value of  the  critical  exponent  in  the
superconducting   correlations and  decrease of those in the  CDW  and
the SDW correlations.

  For the one-dimensional  d-p  model in the weak coupling regime,  T.
Matsunami  and  M.  Kimura\cite{Matsunami}  have  calculated  critical
exponents  in some correlation functions by the renormalization  group
analysis.
They  showed  that the d-p model is mapped onto  $g$-model  with  spin
independent  couplings :
$$
           g_1=U_d|\alpha^+_{k_F}|^4+4U_{pd}\cos{(k_F)}|\alpha^-
_{k_F}|^2|\alpha^+_{k_F}|^2  \eqno{(9.a)}
  $$
 $$
          g_2=                  U_d|\alpha^+_{k_F}|^4+4U_{pd}|\alpha^-
_{k_F}|^2|\alpha^+_{k_F}|^2 \eqno{(9.b)}
$$
  where                           $|\alpha^\pm_{k_F}|^2=\frac1{2}(1\pm
\Delta/\sqrt{\Delta^2+4t^2_k})$.
  The T-L regime  corresponds to the case with $g_1>0$, while the  L-E
regime corresponds to the case with $g_1<0$.
 In  the most divergent approximation, the critical exponent  $K_\rho$
is given by $K_\rho=1+(g_1-2g_2)/(2\pi v_F)$.
The superconducting phase  appears in the region of $g_1-2g_2>0$.
The transformed phase diagram on the $U_d$-$U_{pd}$ plane is shown  in
Fig.1.

\section{ Numerical results}

In  the  non-interacting case,  the critical  exponent  $K_{\rho}$  is
unity  at any $\Delta$ and any filling for infinite systems.
Numerical results of $K_{\rho}$  indicate that $K_{\rho}$ is equal  to
0.92 for the system of 6-unit cells with  8-electrons  and 0.90 for 4-
unit cells with 4-electrons at $\Delta=2$.
It  suggests that the size dependence of $K_{\rho}$  are not so  large
even the 4-units system.

For  finite coupling case, the numerical values of $K_\rho$  are  0.86
(0.91)  ,0.80  (0.81),  and 0.77 (0.72) for $U_d=$0.5,  1.0,  and  1.5
respectively  in  the  system of 6-unite cells   with  8-electrons  at
$\Delta=2$  and  $U_{pd}=0$,  where the  values  in   parentheses  are
obtained by the  g-ology.
The  consistency of both results is good in the weak coupling  regime.
We expect that our numerical results are sufficiently reliable for the
strong coupling regime.

 \subsection{Phase diagram at half-filling}
First we investigate  the electronic state at  half-filling ($n$=1).
 In  Fig.2, we  show the contour map for $K_\rho$ on the  $U_d-U_{pd}$
plane  at $\Delta=2$ for the system of 6-unit cells with  6-electrons,
where the contour lines are drawn by using the spline interpolation.
 The numerical result agrees with the result of $g$-ology in the  weak
coupling regime.
 It  indicates that  the superconducting phase (SC(I)) appears in  the
lower half-plane ( $U_{pd} \displaystyle{ \mathop{<}_{\sim}} 0)$.
 The   attractive   interaction   $U_{pd}$     accounts   for     this
superconducting phase.
  Moreover, in the strong repulsive regime, a novel    superconducting
phase (SC(II)) appears for the large repulsive interaction $U_{pd}$
   in the proximity of the  phase separation (PS(II)).
 These results seem to be similar to that  of  the one  band model( U-
V model) obtained in the previous work\cite{Mila,Sanoono2}.

   We also  evaluate the charge excitation gap $\Delta_{\rm C}$ by the
discontinuity  in the chemical potential  at half-filling ($n=1$)  and
show the region of the  insulating state with finite $\Delta_{\rm  C}$
in Fig.2.
  For large $U_d$ and $U_{pd}$,  we find that the charge gap exists on
a large area.

   In the case of $U_d=\infty$,    we have shown that the charge   gap
$\Delta_{\rm  C}$   is  roughly proportional  to  $\Delta$  for  large
$\Delta$\cite{Sanoono1}.
 By  using the least square method, we estimate $\Delta_{\rm  C}$   in
more  detail   as  a  function of  $\Delta$     for  $U_d=\infty$  and
$U_{pd}$=0 in Fig.3.
   It  indicates  that $\Delta_{\rm C}$ is   roughly  proportional  to
$\Delta-2t^2/\Delta$  for large  values of $\Delta$\cite{cgap}.
 The  effect of $U_{pd}$  is to increase $\Delta_{\rm C}$  by  roughly
$2U_{pd}$,  which  corresponds  to  the  mean  field  contribution  of
$U_{pd}$ to $\Delta_{\rm C}$.
Accounting this  effect, we estimate the critical value $U_{pd}^c$  of
the metal-insulator transition at $\Delta=2$.
 In  this  case,  we get $\Delta_{\rm C}\sim1.3$  for  $U_{pd}=0$  and
$U_{pd}^c=-\Delta_{\rm C}/2\sim-0.7$.    This result consists with the
numerical result in Fig.2.

On  the  other  hand, in the limit $U_{pd} \to  \infty$,  the  problem
becomes very simple and we can get some exact results.
The exact wave function of the insulating  state is given by
 $$ |ins>=|.....,\b1,0,\b1,0,\b1,0,\b1,0,\b1,0,......>,    $$
 where  each  underlined number denotes the number of electrons  at  a
Cu-site and another does the number of electrons at a O-site.
The energy  of this insulating state is $E_{ins}=0$.
We  also  consider  a metallic  state defined by  the  first   charge-
excited state from the insulating state, whose element is presented by
 $$
 |metal>=|....,0,\b1,0,\b2,0,\b h,0,\b1,0,......>,
 $$
 where   "2"  denotes a doubly occupied site ( we call  this  electron
pair   $dimer$ ) and "$h$" denotes a $hole$ which can move  freely  in
the   d-like   band.   The   energy   of   the   metallic   state   is
$E_{metal}=U_d+E^-(k=0)$,  where  $E^-(k)$ is defined by  eq.(2)  $E^-
(k=0)=-1.236..$ in the case of $\Delta=2$.
When $E_{metal}>E_{ins}$, the insulating  state is the ground state.
Then,  the metal-insulator transition occurs at  $U_d^c=-E^-(k=0)$  in
the limit $U_{pd}\to\infty$.
 Our numerical result  agrees with the exact  result  mentioned above.

When  $U_d$  is smaller than $U_d^c$, many  charge  excitation  occur;
which  make  dimers and holes.  To gain the kinetic energy  of  holes,
dimers  are  required to aggregate each other.  Therefore,  the  phase
separation (PS(II)) occurs such as
$$
|\mbox{PS(II)}>=|...,\b2,  0,\b2, 0,\b2, 0,\b2, 0,\b1,  0,\b  h,0,\b1,
0,\b h,0,\b1,0,...>
$$
  as seen in Fig.2.
When  $U_d$  becomes large negative value,  all of the  electron  make
dimers and the phase separation disappears.

 \subsection{Phase diagram  away from half-filling}

Next   we investigate the electronic state away from half-filling.  In
Fig.4,  we  show  the phase diagram on   the  $U_d-U_{pd}$  plane  for
$n=4/3$( 8-electrons per 6-unit cells).

In  the  attractive coupling regime with $U_d<0$  or  $U_{pd}<0$,  the
phase diagram is similar to that of half-filling.
In  the  repulsive  coupling regime with  $U_d>0$  and  $U_{pd}>0$,the
phase  separation  region (PS(III)) appears instead of  the  insulator
phase  seen  in half-filling. Furthermore,  the  new   superconducting
phase (SC(III))  is also  found in the proximate  region  towards  the
phase separation.
  The  mechanism of the phase separation PS(II) can be  understood  as
well as  that of  the half-filling case.
However,  for PS(III), the mechanism of the phase separation   changes
completely,   since double occupancy of electrons on each  Cu-site  is
prohibited  and doped electrons (they make dimers) must sit at O-sites
at the cost of the atomic energy difference $\Delta$.
In  the  limit of $U_{d}$ and $U_{pd} \to \infty$,  a  schematic  wave
function of PS(III)  is given as
$$
   |\mbox{PS(III)}>=|....,\b0,2,\b0,2,\b0,2,\b0,0,\b1,0,\b1,0,\b1,0,....>.
$$
Again, dimers are aggregated each other to gain the kinetic energy. Of
cause,  for  $U_d<2\Delta$,  dimers can sit also on Cu-sites  and  the
above  type  of  the  phase separation vanishes.  It  agree  with  our
numerical result in Fig.4.

 \subsection{ Analysis of Spin gap  }
Finally, we consider  the energy gap in the spin excitation spectrum.
For weak coupling regime, the boundary line  between T-L regime and L-
E regime obtained by  the g-ology on the $U_d-U_{pd}$ plane is   $g_1=
0$, i.e.,
 $  U_d=-4U_{pd}\cos{(k_F)}|\alpha^-_{k_F}|^2/|\alpha^+_{k_F}|^2    $
(See  Fig.1).
For  a fixed value of $U_{pd}$, the spin gap  vanishes at  a  critical
value  $U^s_d$.  In the weak coupling regime, $U^s_d$  increases  with
$U_{pd}$ for $n<1(k_F<\pi/2)$.
On the other hand, in the strong coupling limit  $U_{pd}\to\infty$,the
spin  gap   vanishes at the  critical  value  $U^s_d=2E^-(k=0)=-2.472$
exactly\cite{Upd}.
 To  presume the boundary line  for intermediate coupling regime,   we
will  estimate the spin gap numerically.
 The spin excitation energy for a finite size system is obtained  from
the energy difference between the lowest triplet state and the singlet
ground state.  We assume that the size dependence of the spin gap as
$$
 \Delta_{\rm{S}}(N_u)^2=\Delta_{\rm{S}}(\infty)^2+C/N_u^2,       \quad
\eqno{(10)}
  $$
where $\Delta_{\rm{S}}(N_u)$ is the spin gap of $N_u$-site system  and
$C$  is a  constant. Sano and Takano\cite{Sanotakano} have shown  that
this  finite size scaling is successful in the estimation of the  spin
gap  for $t-J-J'$ model.
Hereafter,    we    denote    the     $\Delta_{\rm{S}}(\infty)$     as
$\Delta_{\rm{S}}$ for simplicity.

We  calculated   the $\Delta_{\rm{S}}$ as a function  of  $U_d$  under
fixed  values  of  $U_{pd}$ using the eq.(8) with the  $N_u=$3  and  6
systems for $n=4/3$.
Fig.5   shows   that   a  large  spin  gap   opens   in   the   region
$U_d\displaystyle{   \mathop{<}_{\sim}   }-2$.  It   decreases    with
increasing   $U_d$  and  seems  to close to  zero  at  $U_d\sim0$  for
$U_{pd}=0$.
We get  $U^s_d\displaystyle{ \mathop{>}_{\sim} }-0.5$ for $U_{pd}=$1,2
and 3 and  $U^s_d\sim-1$ for $U_{pd}=$5.
Because  of the finite size effect, it is difficult to  determine  the
$U^s_d$ accurately.
 The   result  of  the  limiting  case   $U_{pd}\to\infty$   ($U^s_d=-
2.472...$),  the  $g$-olgy and  the numerical data for  $U_{pd}\le  5$
suggest  that  the  superconducting phases SC(I) and SC(II) belong  to
L-E  regime in part, while the  another superconducting phase  SC(III)
belong to L-E regime in whole\cite{spingap}.

 \section{Summary and discussion}
In  summary, we have numerically diagonalized the one-dimensional  d-p
model  with  finite sizes.  Paying special attention to  the  role  of
$U_{pd}$  in  the  strong  coupling regime,  we  have  calculated  the
critical exponent of correlation functions,   charge gap and spin  gap
in the systems.
On  the $U_d-U_{pd}$ plane, the obtained phase diagram  consists  with
the result of g-ology in the weak coupling regime.
On  the  other hand, in the strong coupling regime,  more  complicated
phases  are obtained.
Using    the   Luttinger  liquid  relations,  we   found   the   three
superconducting phases (SC(I),SC(II) and SC(III)) in the proximity  of
the phase separation region ( PS(I),PS(II) and PS(III) ) respectively.
   The  relation  (8) tells us that the   enhancement   of    $\chi_c$
leads  to increase of absolute value of the critical exponent  in  the
superconducting   correlations and  decrease of those in  the CDW  and
the SDW correlations.
Thus, unstable phase is accompanied by the superconducting phase.

We  speculate that the  three types of mechanism of  phase  separation
produce   the three types of superconducting phases.
The origin of unstable phase PS(I) and the superconducting phase SC(I)
are explained by the attractive interaction $U_d$ and/or $U_{pd}$.
The  large repulsion $U_{pd}$ produces the unstable phase  PS(II)  and
the  superconducting phase SC(II) at half-filling (similar phase   has
been obtained in the 1D $U-V$ model)\cite{Mila,Sanoono2}.
  Away  from  half-filling, the third unstable phase PS(III)  and  the
superconducting  phase  SC(III)  appear in  the  region  $U_d>>0$  and
$U_{pd}>0$ .
  Our result suggests  that $U_{pd}$ enhances the charge   fluctuation
which promotes the superconducting correlation.
It  is noted that the parameter region of SC(III) is corresponding  to
the charge transfer insulator region at half-filling. We believe  that
this   result    have   the  relevance    to   the    high-temperature
superconductivity  which is realized by doping to the charge  transfer
insulator.

 The  spin  gap analysis suggests that  the SC(I)  and  SC(II)  phases
partially include the Luther-Emery region (with spin-gap). At the same
time,   the whole SC(III) phase belongs to  Tomonaga-Luttinger  liquid
(without spin-gap).
Generally, in realistic parameter region  $U_d>>U_{pd}>0$, the  system
belongs to  the T-L regime. It is also checked as follows.
 we investigate the finite size correction to the ground state  energy
$E_0$.  The  conformal filed theory predicted the size  dependence  of
$E_0$ as
$$
E_0/N_u =\epsilon-\frac{\pi}{6} \frac{c(v_c+v_s)}{N_u^2}+o(N_u^{-4}),
  \eqno{(11)}
$$
 where  $v_s$ is the spin velocity and $c$ is the central  charge.  If
the system is a Luttinger liquid, we should have $c$=1.
 This was shown explicitly for 1D Hubbard model as well as for the 1D-
supersymmetric $t-J$ model.
  For 1D  $d-p$ model, we calculate the central charge numerically  by
fitting  the ground state energy to the formula of the above  equation
by taking $n=$4/3 with the system size $N_u$=3, 6.
The  charge and spin velocity, $v_c$ and $v_s$ are estimated  for  the
largest  6-unit  cluster. It turned out that $c$  is  almost  constant
$(0.9<c<1.1)$  in the region $U_{pd}=0.0\sim11.5$ for  $\Delta=4$  and
$U_{d}=\infty$.  It  suggests that the 1D $d-p$ model behaves  as  the
Tomonaga-Luttinger liquid for $U_d>>0$.

\section*{Acknowledgement}
    This  work is partly supported by the Grant-in-Aid for  Scientific
Research on Priority Areas, Mechanism of Superconductivity  (02213103)
from the Ministry of Education, Science and Culture.

\clearpage
\noindent
{\large \bf Figure Captions}

\begin{description}

\item[Fig.1]
The phase diagram  on the $U_d-U_{pd}$ plane obtained by the  g-ology,
where $g_1$ is given by eq.(10.a) and $g_1-2g_2=
  -(U_d|\alpha^+_{k_F}|^4+4U_{pd}|\alpha^-
_{k_F}|^2|\alpha^+_{k_F}|^2(2-\cos{(k_F)})).    $
Note  that,  in the repulsive regime $U_d>0$  and  $U_{pd}>0$,  $(g_1-
2g_2)$ is always negative and then the superconducting phase does  not
appear within the $g$-olgy.

\item[Fig.2]  Contour map for the $K_\rho$ on the  $U_d-U_{pd}$  plane
at  $\Delta=2$.  We  used the 6-unit  6-electron  system  ($n$=1)  and
calculated the values of $K_{\rho}$ at points of $U_d=-5, -4,...., 10$
for  $U_{pd}=-2, -1,..., 10$.
 The contour lines are plotted by using the spline interpolation.

\item[Fig.3]  The   charge  gap $\Delta_{\rm{C}}$  as  a  function  of
$\Delta$  for  $n=1$. A dotted line  represents  $\Delta-2t^2/\Delta$,
which   is  obtained  the  second  order  perturbation   method   (See
Ref.\cite{cgap}).

\item[Fig.4]  Contour map for the $K_\rho$ on the  $U_d-U_{pd}$  plane
for $n$=4/3 at $\Delta=2$. We used the 6-unit  8-electron system.

\item[Fig.5]
 The  spin  gap  $\Delta_{\rm{S}}$    as  a  function  of  $U_d$   for
$U_{pd}$=0,1,2,3 and 50 at $n=$4/3.

\end{description}


\begin{thebibliography}{99}
\bibitem{cgap} The charge gap $\Delta_{\rm{C}}$ is calculated by   the
second-order perturbation with respect to $t$  for large $\Delta$.  We
choose   the   initial   state   $|...,0,\b1,0,\b1,0,\b1,0,...>$   and
intermediate    state    of    the    charge    transfer    type    as
$|....,,0,\b1,0,\b0,1,\b1,0,...>$.
\bibitem{Upd}
In  the  limit $U_{pd}\to\infty$, a ground state  for  large  negative
$U_{d}$ is given as

$$ |G. S>=|...,0,\b2,0,\b2,0,\b2,0,\b0,0,\b0,0,\b0,0,\b2,0,....>.$$
In  this case, the kinetic energy of electrons is zero and the  ground
state  energy $E_g$ is $(N_u/2)U_{d}$ where we set $\epsilon_d=0$.  If
the charge excited occur, a dimer brakes up and two electrons can move
freely.  Thus,  the  energy $E_{ex}$ of this    excitation  state   is
$(N_u/2-1)U_{d}-2E_-(k=0)$.  When $U_{d}=-2E_-(k=0)$,  the  excitation
energy is zero and the spin gap vanishes.
\bibitem{spingap}
For $n=1$, the $g$-ology shows the boundary line ($g_1=0$) is given by
$U_d=0$.
Numerical  data of $\Delta_{\rm{S}}$  seem to suggest that   the  line
borders  the left of the boundary of the superconducting phase  SC(II)
in Fig.2.
\end{thebibliography}
\end{document}